\documentclass[11pt]{amsart}
\usepackage{studia}
\usepackage[table]{xcolor}
\usepackage{graphicx}

\coordinates{LVII}{1}{2012}

\title{A Software Repository and Toolset for Empirical Research}
\author{Arthur-Jozsef Molnar}
\address{Department of Computer Science, Faculty of Mathematics and Computer Science,Babe\c{s}-Bolyai University, 1, M. Kogalniceanu, Cluj-Napoca 400084, Romania}
\email{arthur@cs.ubbcluj.ro}
\date{31.01.2012} 


\subjclassCR{%
D.2.0 [\textbf{Software}]: Software Engineering -- \textit{General};
}


\begin{document}

\begin{abstract}
This paper proposes a software repository model together with associated tooling and consists of several complex, open-source GUI driven applications ready to be used in empirical software research. We start by providing the rationale for our repository and criteria that guided us in searching for suitable applications. We detail the model of the repository together with associated artifacts and supportive tooling. We detail current applications in the repository together with ways in which it can be further extended. Finally we provide examples of how our repository facilitates research in software visualization and testing.
\end{abstract}

\maketitle

\section{Introduction}

Empirical methods are of utmost importance in research, as they allow for the validation of formal models with real-life data. Many recent papers in software research contain sections dedicated to empirical investigation. In \cite{189}, Kitchenham et al. propose a set of guidelines for researchers undertaking empirical investigation to help with setting up and analyzing the results of their investigation. Stol et al. survey existing empirical studies concerning open source software in \cite{190}, presenting predominant target applications together with employed research methods and directions. More recently, Weyuker overviewed the development of empirical software engineering, highlighting notable success stories and describing current problems in the field \cite{191}.

While software literature abounds with empirical research \cite{191}, we find a lack of advanced tooling to support such endeavours. Many of the required steps in carrying out empirical research are repetitive in nature: find suitable applications, set them up appropriately, configure them, run the desired experimental procedures and conclude the procedure. Our aim is to assist researchers interested in carrying out empirical investigation by providing an extensible repository of complex open-source software that is already set up and has multiple associated software artifacts that simplify research in areas such as code analysis, software visualization and testing.

An important step was to study existing research-oriented repositories in order to take necessary steps to address shortcomings of previous approaches. Herraiz et al. \cite{188} describe approaches to obtain research datasets from freely available software repositories such as SourceForge\footnote{http://sourceforge.net/} together with associated mining tooling such as FLOSSmole\footnote{http://flossmole.org/} \cite{192}. In \cite{187} Alexander et al. present the \emph{"Software Engineering Research Repository (SERR)"} \cite{187} which consists of source code, software models and testing artifacts that are organized by project. Authors describe SERR as a distributed repository that is easy to extend and which can bring benefits related to software education, research and to provide solid artifact examples for developers.

This paper is structured as follows: the following section details our criteria in searching for suitable applications, while the third section details academic tools used to obtain some of the complex artifacts in our repository. The fourth section is dedicated to detailing our repository's model while the fifth section details its contents. Previous work using the repository is detailed in the sixth section while the last section presents conclusions and future work.

\section{Criteria for Suitable Applications}

In this section we detail criteria that guided our search for applications suitable as targets for empirical research in domains such as software visualization, analysis and testing. We believe the following criteria to be generally useable in assessing the fitness of candidate software applications as targets of empirical research:
\newline
\begin{itemize}
\item \emph{Usability}: We believe all meaningful research must target useful software. While a broad term, our opinion is that a vibrant user and developer community that share ideas and actively guide the development of the application to be a good sign of such software. Also, many hosting sites such as SourceForge provide detailed download statistics that can be used to assess real life usage for candidate applications. We present relevant statistics for applications in our repository in the 5th Section of the present paper.
\newline

\item \emph{Complexity}: Real software is complex and today we have lots of methodologies and metrics for assessing software complexity. Thorough research must be accompanied by several relevant complexity metrics and should be tied with the usability criteria mentioned above.
\newline

\item \emph{Authorship}: We believe that picking applications totally unrelated to the research effort goes a long way in proving its general applicability. The best way to achieve this is to select target applications produced by third parties that are uninterested and unrelated to the undertaken research efforts \cite{189}.
\newline

\item \emph{Availability}: The first requirement for using software is of course its availability. This has multiple implications that are best described separately:
\newline

\begin{itemize}
\item \emph{Legal}: It is best when target applications are provided under a free software style license. This allows unrestricted access to view or alter source code as needed and the possibility of further distribution within the academic community for purposes of validating research or allowing others access to expand it. In this regard free software licenses\footnote{http://www.gnu.org/licenses/license-list.html\#GPLCompatibleLicenses} compel those who alter the target software to continue providing the modified versions under the same license, preventing legal lock-in.
\newline
\item \emph{Technical}: Many research efforts require tracking the evolution of the applications under study. For example, Memon uses a total of 19 software versions of FreeMind \cite{103}, GanttProject\footnote{http://www.ganttproject.biz} and JMSN\footnote{https://sourceforge.net/projects/jmsn} in a case study that attempts to repair GUI test cases for regression testing \cite{97}. In \cite{67}, Xie and Memon present a model-based testing approach where a total of 24 versions of four open-source applications are used as case study targets. In addition, our research requires access to multiple target application versions. Our jSET visualization and analysis tool \cite{180} was tested using multiple versions of applications in our repository, which were again reused in our work in heuristically matching equivalent GUI elements \cite{177}. All these efforts were possible by having open-source applications available that have a public source code repository such as SourceForge with available change history going back several years.
\end{itemize}
\mbox{}
\item \emph{Simplicity}: The need to access multiple versions of the same software application raises the importance of the simplicity criterion. In order to have multiple programs set up, configured and ready to run applications should not require complex 3rd party packages such as relational database management systems or web servers that many times are challenging to set up, configure and clean up once investigation has concluded. This has become apparent both in our research \cite{177,180} and in case studies by Xie and Memon \cite{67,97}, Xiao et al. \cite{182}, Do and Rothermel \cite{183,184} and many others.
\end{itemize}

\section{Related Tools}

When developing our repository we made a goal out of harnessing advanced research frameworks to obtain complex artifacts associated with the included applications. In this section we briefly describe the Soot\footnote{http://www.sable.mcgill.ca/soot} static analysis framework and the GUITAR\footnote{http://sourceforge.net/apps/mediawiki/guitar/index.php?title=GUITAR\_Home\_Page} testing framework which we use to obtain complex artifacts that describe our repository applications.

Soot is a static\footnote{It does not run the target application} analysis framework targeting Java bytecode \cite{42,44,46}. Currently there are many types of analyses Soot can perform \cite{44,46}, some of which are planned for future integration into our repository. One of the most important artifacts produced by Soot is the application's call graph: a directed graph that describes the calling relations between the target application's methods \cite{44}. The graph's vertices represent methods while the edges model the calling relations between them. Being computed statically, it does not provide information regarding the order methods are called or execution traces. This static callgraph is an over-approximation of all the dynamic callgraphs obtained by running the application over all its possible inputs. Of course, this means the graph will contain spurious edges and nodes, some of which can be eliminated by using mode advanced algorithms \cite{46}. By default, we use the SPARK engine detailed in \cite{44} to obtain callgraphs for applications in our repository.

The second tool we employ is the GUIRipper application part of the comprehensive GUITAR toolset \cite{142}. GUIRipper acts on a GUI driven target application \cite{20} that it runs and records all the widgets' properties across all application windows. It starts the target application and records the properties of all the widgets present on the starting windows and fires events on them\footnote{Clicking buttons, selecting menu items} with the purpose of opening other application windows that are then recorded in turn. The resulting GUI model described in \cite{20} is then persisted in XML format for later use. It is important to note that the only required artifact is the target application in compiled form. Although completely automated, GUIRipper's behaviour can be customized using configuration files. This makes it possible to avoid firing events with unwanted results, such as creating network connections, printing documents and so on. The GUIRipper tool is available in versions that work with Microsoft Windows and Java applications \cite{102}. We currently use the Java implementation of the tool to capture GUI models for applications in our repository.

The following section presents our proposed repository model and details our changes to Soot and GUITAR that allow recording more information about target applications.

\section{The Repository Model}

Our repository is modeled as a collection of \emph{Projects}, where each \emph{Project} represents a target application at a given moment in time. For example, a project describes the FreeMind application detailed in the 5th Section as found on its SourceForge CVS on November 1st, 2000. Other projects can represent the same application at different points in time. Having multiple projects for the same application helps when studying regression testing, tracking and analyzing software evolution and much more. Each project consists of the following artifacts:

\begin{itemize}
\item \emph{Project File}. This is an XML file that contains the project's name and location of associated artifacts. It is similar in function to Eclipse's\footnote{http://eclipse.org/} \emph{.project} and Visual Studio's\footnote{http://msdn.microsoft.com/en-us/vstudio/aa718325} \emph{.sln} files.
\item \emph{Application binaries}. Each project contains two sets of binary files: the compiled application and its required libraries. We currently use different directories for each of these sets of files. Also, the directory containing application binaries contains a script that starts the application. While this may appear trivial, it becomes important when multiple versions of the same application are present. Many times applications record user options and configurations in system-specific locations (e.g: \%WINDOWS\%/Users) which if not properly handled might cause other versions of the same application to malfunction or provide inconsistent behaviour\footnote{Especially as many applications were not designed to co-exist in multiple versions on the same machine}. To mitigate these aspects we manually checked each application version's source code and created adequate startup scripts to set up a consistent environment for the application, which is cleaned after the program's execution.
\item \emph{Application sources}. Each project has a source directory that contains the application's source code. The sources should be complete in the sense to allow recompiling the application.
\item \emph{GUI Model}. Contains the XML model obtained by running our modified version of GUITAR's GUIRipper on the target application.
\item \emph{Widget Screenshots}. Our modified version of the GUIRipper application records screenshots that are associated with the GUI widgets. This allows studying the widgets' appearance without having to start the application. Due to their large number, screenshots are not stored on our SVN repository \cite{105}, but are easy to obtain by running the scripts that record the GUI model. This way they will be automatically placed in the correct project subfolder.
\item \emph{Application callgraph}. This is the callgraph obtained by running our Soot wrapper over the target application.
\end{itemize}

Our repository \cite{105} is structured so that every project has its own SVN directory, making it easy to select and download individual projects. In addition to providing actual repository data, our toolset implemented using the Java platform allows programmatic access to target application data. Each project is programmatically represented by an instance of the \emph{jset.project.Project} class which provides access to the project artifacts discussed above. Loading \emph{Project} instances is done via the \emph{jset.project.ProjectService} class that provides the required methods. Our model provides access to method bytecode using the BCEL\footnote{Bytecode Engineering Library - http://commons.apache.org/bcel} library that we use to parse compiled code as shown in Figure \ref{fig:CallgraphModel}. More so, loaded \emph{jset.project.Project} instances link method bytecode with available sources using  Eclipse JDT \cite{111} as a source code parser.

For handling complex artifacts, our repository projects contain all necessary scripts together with Soot and GUIRipper configuration files that allow recomputing the callgraph and GUI model at any time. A lot of work was dispensed when building the required configuration files to make sure all aspects of the target applications are correctly recorded in a repeatable manner. This effort was also reflected in the scripts used when starting the application to make sure that application GUIs remain consistent across multiple executions.

The following sections further detail our model by presenting our changes to the GUITAR and Soot frameworks together with our GUI and callgraph model implementations. Also, we describe how projects can be obtained in an automated build environment.

\subsection{GUI model}
\label{sub:OurGUIModel}

To make the most of available tooling, we started by modifying GUITAR's GUIRipper to provide some additional model information: 

\begin{itemize}
\item \emph{Widget screenshots}. Every time the GUIRipper records a GUI element it will also save a screenshot of its parent window. The element's associated screenshot together with location and size on the containing window are recorded among its properties so it can be later located by our  software tools.
\item \emph{Event handlers}. Our modified version of GUIRipper records all event handlers of recorded GUI elements. This provides the link between the GUI layer and its underlying code.
\end{itemize}

The GUI model is persisted in GUIRipper format, but on programmatically loading the project it is converted into a simpler model we developed to gain independence from external implementations. While adding an extra step, our model provides simplicity while allowing better control when using other model sources such as XAML \cite{157}, UIML \cite{79} or HTML. This becomes important as extending our repository beyond the desktop paradigm is a target for future work. Also, by persisting the model in its original format makes it easy to compare models without first having to convert them, which is important when setting up the process for a new target application. 

\begin{figure}[h]
	\centering
	\includegraphics[scale=1]{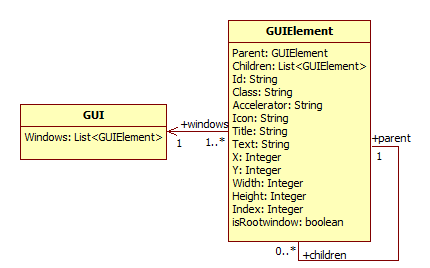}
	\caption{GUI model}
	\label{fig:JSET_GUI_Model}
\end{figure}

Our hierarchical GUI model is shown in Figure \ref{fig:JSET_GUI_Model}. Each GUI is represented using an instance of the \emph{GUI} class, which has no physical correspondent on the actual user interface. Its windows and widgets are represented by \emph{GUIElement} descendants as such: windows are the first level children of the root node while contained widgets descend from their containing windows true to the GUI being modeled. Most GUI element attributes are represented with simple data types such as String or Integer. An important aspect regards the \emph{Id} property which is used by our process to uniquely identify a graphical element. The \emph{Id} must be provided externally to our process as a String instance with the restriction that each element must have a unique identifier. This is enforced in our implementation using safeguard checks that take place when loading the project. 

Externally obtained models are converted into our own by providing an implementation for the \emph{IGUIModelTransformer} interface. One such implementation is already available for use with GUIRipper models; other model sources require implementing a suitable transformer.

\subsection{Callgraph Model}

\begin{figure}[h]
	\centering
	\includegraphics[width=1\textwidth]{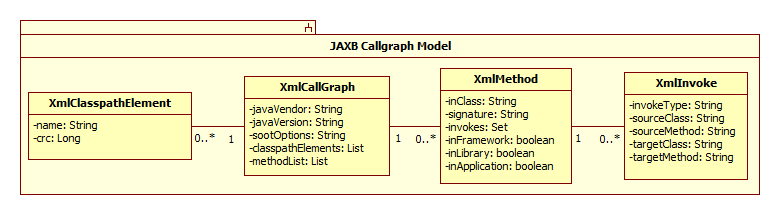}
	\caption{JAXB-backed callgraph model}
	\label{fig:JAXBCallgraph}
\end{figure}

Soot does not provide an endorsed model to persist computed callgraphs so we implemented a wrapper over Soot that persists the callgraph using a JAXB-backed model detailed in Figure \ref{fig:JAXBCallgraph}. The central class is \emph{XmlCallGraph} which holds a collection of \emph{XmlMethod} instances. Note that the callgraph model does not contain classes directly, but they can be inferred using the \emph{inClass} attribute of \emph{XmlMethod} instances.

An important aspect regards the three boolean attributes of the \emph{XmlMethod} class: \emph{inFramework}, \emph{inLibrary} and \emph{inApplication}. When computing the application call graph, we divide analyzed classes into framework, library and application. Framework classes are ones provided by the Java platform itself in libraries such as \emph{rt.jar}, \emph{jce.jar}\footnote{On the Oracle Java implementation} and so on. Library classes are usually provided on the classpath while application classes are the ones that actually comprise the target software. As the callgraph only models methods and not classes, this information is persisted using them. Also, we must note that multiple classes with the same name might be present in a virtual machine\footnote{The Xerces XML library provides such an example}, so the three categories are not mutually exclusive.

\begin{figure}[h]
	\centering
	\includegraphics[width=1\textwidth]{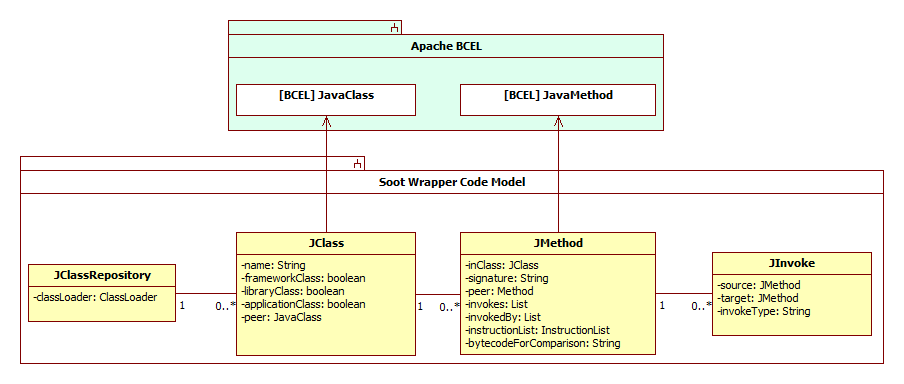}
	\caption{Project code model}
	\label{fig:CallgraphModel}
\end{figure}

When a \emph{Project} instance is loaded, the XML model is read and callgraph information is combined with static class data obtained using the BCEL\footnote{http://commons.apache.org/bcel/} library. The obtained model is shown in Figure \ref{fig:CallgraphModel}. \emph{JClass} and \emph{JMethod} instances are wrappers around BCEL implementations and provide class and method level details such as line number tables, constant pools and method instruction lists in human readable form. The model is browsable using methods provided in \emph{JClassRepository}.

Our model's most pressing limitation concerns obtaining the project's code model shown in Figure \ref{fig:CallgraphModel}. As BCEL and Soot only work on Java programs, the model cannot be constructed for applications implemented using other platforms. While specialized tools equivalent to BCEL exist for other platforms, we do not have knowledge of tools equivalent to Soot. This leads to the limitation that complete \emph{Project}s can only be recorded for Java software.

\subsection{Automatically Building Projects}

Because each project instance captures a snapshot of the target application, it raises the interesting question of integrating project building with application development. This can help with setting up multiple software versions for research automatically or for purposes of software visualization and analysis \cite{177,180}. Therefore one of our goals was to provide means to enable automatically building projects from target application sources. Application sources and associated binaries can be obtained using a regular nightly/weekly build process. Obtaining secondary artifacts such as the GUI model and application callgraph is possible using scripts such as ones provided with all the projects currently in our repository. 

Automating the process for a new target application requires the creation of suitable script files that compile the application, run our modified version of GUIRipper and Soot wrapper over the compiled sources and place the obtained artifacts into proper directories. Both GUIRipper and our Soot wrapper are extensively customizable using configuration files which only need to be updated in case of major GUI or application classpath changes. 


\section{Repository Contents}
\label{sub:RepositoryContents}

The search based on the criteria laid out in the 2nd section led us to two target applications: the FreeMind \cite{103} mindmapping software and the jEdit \cite{104} text editor. Both of them are available on the SourceForge website and are provided with free-software licenses that allow altering and distributing their source code. The present section discusses both these applications.

\subsection{FreeMind}
\label{sec:FreeMind}

The FreeMind mind mapping software is developed on the Java platform and is available under GNU's GPL license on Windows, Linux and Mac platforms. Our repository contains 13 distinct versions of the FreeMind application, dated between November 2000 and September 2007. We used a script to download weekly snapshots of the application starting with the earliest stable version, released as 0.2.0. The development process was not steadily reflected in the source code repository as we found a notable hiatus in CVS data between 2005 and 2006. Out of the weekly versions obtained we found most to be identical regarding source code. Using manual examination we distilled the available data to 13 versions that have differences in source code. Table \ref{tab:FreeMindVersionHistory} contains details regarding the downloaded versions such as CVS time, approximate corresponding release version and the number of classes, lines of code and GUI widgets recorded.

\begin{table}[h]
\centering
\begin{tabular}{ | c | c | c | c | c | c | }
 \hline
\rowcolor{lightgray} Version & CVS Timestamp & Classes & LOC & Widgets & Windows\\ \hline
0.1.0 & 01.11.2000 & 77 & 3597 & 101 & 1 \\ \hline
0.2.0 & 01.12.2000 & 90 & 4101 & 106 & 1 \\ \hline
0.2.0 & 01.01.2001 & 106 & 4453 & 132 & 1 \\ \hline
0.3.1 & 01.04.2001 & 117 & 6608 & 127 & 1 \\ \hline
0.3.1 & 01.05.2001 & 121 & 7255 & 134 & 1 \\ \hline
0.3.1 & 01.06.2001 & 126 & 7502 & 136 & 1 \\ \hline
0.3.1 & 01.07.2001 & 127 & 7698 & 137 & 1 \\ \hline
0.4.0 & 01.08.2001 & 127 & 7708 & 137 & 1 \\ \hline
0.6.7 & 01.12.2003 & 175 & 11981 & 244 & 1 \\ \hline
0.6.7 & 01.01.2004 & 180 & 12302 & 251 & 1 \\ \hline
0.6.7 & 01.02.2004 & 182 & 12619 & 251 & 1 \\ \hline
0.6.7 & 01.03.2004 & 182 & 12651 & 251 & 1 \\ \hline
0.8.0 & 01.09.2007 & 544 & 65616 & 280 & 1 \\ \hline
\end{tabular}
\caption{Versions of FreeMind used}
\label{tab:FreeMindVersionHistory}
\end{table}

Regarding the data shown in Table \ref{tab:FreeMindVersionHistory}, some clarifications are in order:

\begin{itemize}
\item The number of classes\footnote{All the metrics were computed using the Eclipse Metrics plugin http://metrics.sourceforge.net} includes those generated from XML models, but does not include library or third-party classes.
\item The number of lines of code (LOC) includes all non-empty and non-comment lines of code. It therefore includes class and class member declarations and other code that in some cases might be considered as overhead.
\item The number of widgets includes all visible or hidden user interface elements recorded by GUIRipper. Therefore this number includes transparent panels employed for grouping controls and elements that might be too small to notice (e.g: 1x1 pixel size) or hidden by Z-ordering.
\item Version number is approximative, as sources were downloaded directly from CVS.
\end{itemize}

With regards to the criteria described in the 2nd section, the FreeMind project was chosen "Project of the Month" in February 2006. The week ending January 29th 2012 saw almost 40.000 downloads, with the total number of downloads being over 14.3 million\footnote{http://sourceforge.net/projects/freemind/files/stats/timeline}.The current version\footnote{as of January 30th, 2012} of the application is 0.9.0 and checking its source control reveals it to be in active development.

By analyzing data in Table \ref{tab:FreeMindVersionHistory} certain interesting facts about the application can be noted. First of all we notice the proliferation of Java classes used to build FreeMind: while the first version in our repository only has 77 classes, most later versions have well above 150 classes, topped by the complex 0.8.0 version with over 500 classes. Also, we witness source code line count being in close progression with the number of classes. An interesting aspect regards the widget count, that increases by a factor of 2.8 during an 18 fold increase in source line count. Another aspect we must mention is that the only window ripped for FreeMind is its main window. Starting with version 0.6.7, FreeMind also has an \emph{Options} window that could not be correctly ripped and so was left out. While not necessarily a threat to the validity of undertaken research, this  aspect must be properly taken into account.

\subsection{jEdit}
\label{sec:jEdit}

The jEdit application is a basic text editor written using Java and similar to FreeMind, is available under the GNU GPL license on multiple platforms. For building our application repository we used a number of 17 versions of this application. While having a public source code repository we chose to select among the many available releases of jEdit for picking the versions. Similar to our approach with FreeMind, we only selected distinct versions (so there are guaranteed code changes between any two versions) that had at least one month of development between them. This allowed us to build a repository containing a reasonable number of applications spread over a number of years with the possibility of including other intermediary versions when required. 

It is worth noting that while FreeMind versions were downloaded directly from CVS, for jEdit we used versions that were released by developers and publicly available on the project's download page. This aspect should be taken into account by users of our repository, as it is not unreasonable to assume that FreeMind versions might display more errors or inconsistencies.

The first version of jEdit considered is the 2.3pre2 version available since January 29th, 2000, while the latest version we used is 4.3.2final, released on May 10th, 2010. As with FreeMind, there was a significant hiatus in the development of jEdit between versions 4.2.0final and 4.3.0final respectively. Table \ref{tab:jEditVersionHistory} presents the versions in our repository together with some key information related to each version, as in the case of the FreeMind application.

\begin{table}[h]
\centering
\begin{tabular}{ | c | c | c | c | c | c | }
 \hline
\rowcolor{lightgray} Version & CVS Timestamp & Classes & LOC & Widgets & Windows \\ \hline
2.3pre2 & 29.01.2000 & 332 & 23709 & 482 & 12 \\ \hline
2.3final & 11.03.2000 & 347 & 25260 & 533 & 14 \\ \hline
2.4final & 23.04.2000 & 357 & 25951 & 559 & 14 \\ \hline
2.5pre5 & 05.06.2000 & 416 & 30949 & 699 & 16 \\ \hline
2.5final & 08.07.2000 & 418 & 31085 & 701 & 16 \\ \hline
2.6pre7 & 23.09.2000 & 456 & 35020 & 591 & 12 \\ \hline
2.6final & 04.11.2000 & 458 & 35544 & 600 & 12 \\ \hline
3.0final & 25.12.2000 & 352 & 44712 & 584 & 13 \\ \hline
3.1pre1 & 10.02.2001 & 361 & 45958 & 590 & 13 \\ \hline
3.1pre3 & 11.03.2001 & 361 & 46165 & 596 & 13 \\ \hline
3.1final & 22.04.2001 & 373 & 47136 & 648 & 13 \\ \hline
3.2final & 29.08.2001 & 430 & 53735 & 666 & 12 \\ \hline
4.0final & 12.04.2002 & 504 & 61918 & 736 & 13 \\ \hline
4.2pre2 & 30.05.2003 & 612 & 72759 & 772 & 13 \\ \hline
4.2final & 01.12.2004 & 650 & 81755 & 860 & 14 \\ \hline
4.3.0final & 23.12.2009 & 872 & 106398 & 992 & 16 \\ \hline
4.3.2final & 10.05.2010 & 872 & 106510 & 992 & 16 \\ \hline
\end{tabular}
\caption{Versions of jEdit used}
\label{tab:jEditVersionHistory}
\end{table}

Please note that the clarifications detailed in the FreeMind section also apply to the present data, excepting the one referring to software versions.

Studying Table \ref{tab:jEditVersionHistory} some interesting facts come to light. First of all we can see the recorded metrics changing as jEdit evolves to a more mature version, almost tripling its number of classes and quadrupling the source code line count. Unlike FreeMind, jEdit presents a user interface that consists of more that 10 windows in each of the recorded versions. However while the code metrics increased significantly, the number of windows did not, staying between 12 and 16 across all the versions (although they become more complex by containing an ever increasing number of widgets). Unfortunately, as in the case of FreeMind, jEdit's \emph{Options} window could not be ripped and was therefore excluded from the data above.

Regarding the eligibility criteria described in the 2nd section, SourceForge reports over 9.000 downloads during the week ending January 29th 2012 and a total number of over 6.7 million downloads since the project was started at the end of 1999. Also, jEdit was selected as SourceForge "Project of the Month" in October, 2010. Its long development history and available download statistics clearly show that jEdit has a large userbase and is in active development with multiple source code commits in January 2012.

\section{Uses of Our Repository}

Our repository is already employed in ongoing research regarding topics such as software visualization, program analysis and automated testing. Its first use was to provide target application input data for the software components in our jSET \cite{116,180} visualization and analysis tool. The multiple application versions were put to good use in our research regarding regression testing of GUI applications \cite{177}, an ongoing effort that will greatly benefit from extending our repository. Having repository data at hand freed us from tasks such as setting up target applications and recording the various required artifacts. Also, the large volume of data allowed carrying out detailed studies and drawing compelling conclusions.

Because both applications that form the bulk of our repository data were used in many previous empirical investigations \cite{99,67,118,194,119,101} we believe our repository model and data are relevant for many future research efforts.

\section{Conclusions and Future Work}

In this paper we presented a proposed software repository model populated with 30 versions of two complex, widely used GUI-driven open-source applications that is freely available for researchers \cite{105}. We detailed changes to popular academic tools that allow recording of key application artifacts together with a software model that allows easy programmatic access to repository \emph{Project} instances. We already employed the presented applications in our research regarding software visualization and analysis \cite{180} and in regression testing GUI applications \cite{177}.

As with many repositories, our first and foremost goal lays in extending it. The main limitation of our repository lays in the fact that both included applications are Java-based. The first future direction is to search for and include software based on other platforms such as .NET or Python. Such efforts must be mirrored by finding suitable applications for code analysis that work on the targeted platform and provide similar functionality to BCEL and Soot, allowing us to build suitable application models.

More generally, we must study how to extend our repository model beyond the desktop paradigm so web and mobile applications can be included. Additional research must be carried out on artifacts relevant for other paradigms together with available tools that can be employed to obtain them. 

A more distant idea is to switch the repository to a distributed model that simplifies future contribution, thus fueling its growth, like proposed in \cite{187}.

\section*{Acknowledgements}
The author was supported by programs co-financed by The Sectoral Operational Programme Human Resources Development, Contract POS DRU 6/1.5/S/3 - ``Doctoral studies: through science towards society''

\bibliography{biblio}
\bibliographystyle{acm}

\end{document}